\documentclass[12pt]{iopart}
\pdfoutput=1
\usepackage{iopams}
\usepackage{amstext}
\usepackage{esint} 
\usepackage{graphicx,color}
\usepackage{array}
\usepackage{setstack}
\usepackage{verbatim}
\usepackage{amssymb}

\usepackage[linkcolor=black,colorlinks=true,citecolor=red]{hyperref} 
\usepackage{etoolbox}

\usepackage{color}

\makeatletter
\def\@mkboth#1#2{}
\newlength\appendixwidth
\preto\appendix{\addtocontents{toc}{\protect\patchl@section}}
\newcommand{\patchl@section}{%
  \settowidth{\appendixwidth}{\textbf{Appendix }}%
  \addtolength{\appendixwidth}{1.5em}%
  \patchcmd{\l@section}{1.5em}{\appendixwidth}{}{\ddt}%
}
\makeatother

\begin{document}

\newcommand{\be}{\begin{equation}}
\newcommand{\ee}{\end{equation}}
\newcommand{\barr}{\begin{eqnarray}}
\newcommand{\earr}{\end{eqnarray}}

\newtheorem{thm}{Main Result}
\newtheorem{lem}{Lemma}
\newtheorem{problem}{Problem}
\renewcommand\thethm{}
\renewcommand\thelem{}
\renewcommand\theproblem{}
\newtheorem{rmk}{Remark}

\newtheorem{claim}{Claim}
 \newcommand{\<}{\langle}
\renewcommand{\>}{\rangle}
\newcommand{\EE}{\mathcal{E}}
\newcommand{\R}{\mathbb{R}}
\newcommand{\C}{\mathbb{C}}
\newcommand{\D}{\mathbb{D}}
\newcommand{\de}{\mathrm{d}}
\newcommand{\XX}{\mathcal{X}}
\newcommand{\BBB}{\mathcal{B}}
\renewcommand{\rho}{\varrho}
\newcommand{\eps}{\epsilon}
\newcommand{\Prob}{\mathrm{Pr}}

\newenvironment{sistema}%
{\left\lbrace\begin{array}{@{}l@{}}}%
{\end{array}\right.}


\title[Universality of the third-order phase transition in the constrained Coulomb gas]{Universality of the third-order phase transition in the constrained Coulomb gas}

\author{Fabio Deelan Cunden$^{1}$, Paolo Facchi$^{2,3}$, Marilena Ligab\`o$^{4}$ and Pierpaolo Vivo$^{5}$}
\address{$1.$ School of Mathematics, University of Bristol, University Walk, Bristol BS8 1TW, United Kingdom}
\address{$2.$ Dipartimento di Fisica and MECENAS, Universit\`a di Bari, I-70126 Bari, Italy\\
$3.$ Istituto Nazionale di Fisica Nucleare (INFN), Sezione di Bari, I-70126 Bari, Italy}
\address{$4.$ Dipartimento di Matematica, Universit\`a di Bari, I-70125 Bari, Italy}
\address{$5.$ King's College London, Department of Mathematics, Strand, London WC2R 2LS, United Kingdom}

\date{\today}

\begin{abstract} 
The free energy at zero temperature of Coulomb gas systems in generic dimension is considered as a function of a volume constraint. The transition between the `pulled' and the `pushed' phases is characterised as a third-order phase transition, in all dimensions and for a rather large class of isotropic potentials. This suggests that the critical behaviour of the free energy at the `pulled-to-pushed' transition may be universal, i.e., to some extent independent of the dimension and the details of the pairwise interaction. 
\end{abstract}

\tableofcontents
\maketitle

\section{Introduction}\label{sec:intro}
In this paper, we are interested in some aspects of the mean field approximation in the statistical mechanics of systems with long-range interactions. 
For these systems, it is possible to characterise the ground state in the thermodynamic limit and compute the (free) energy at zero temperature. In the simultaneous limit of large number of particles and zero temperature, the system typically concentrates in a bounded region $\Gamma\subset\R^n$ and the free energy per particle  attains a finite value. In presence of volume constraints---i.e. if the system is forced within a specified region of space---the ground state and the free energy may be altered. Since phase transitions are associated with the breakdown of the analyticity of thermodynamic potentials, it is legitimate to ask whether the dependence of the free energy on a volume constraint is analytic or not. If not, is it possible to establish what the regularity of the free energy---i.e., the order of the phase transition---is?

\begin{figure}[t]
\centering
\includegraphics[width=.85\columnwidth]{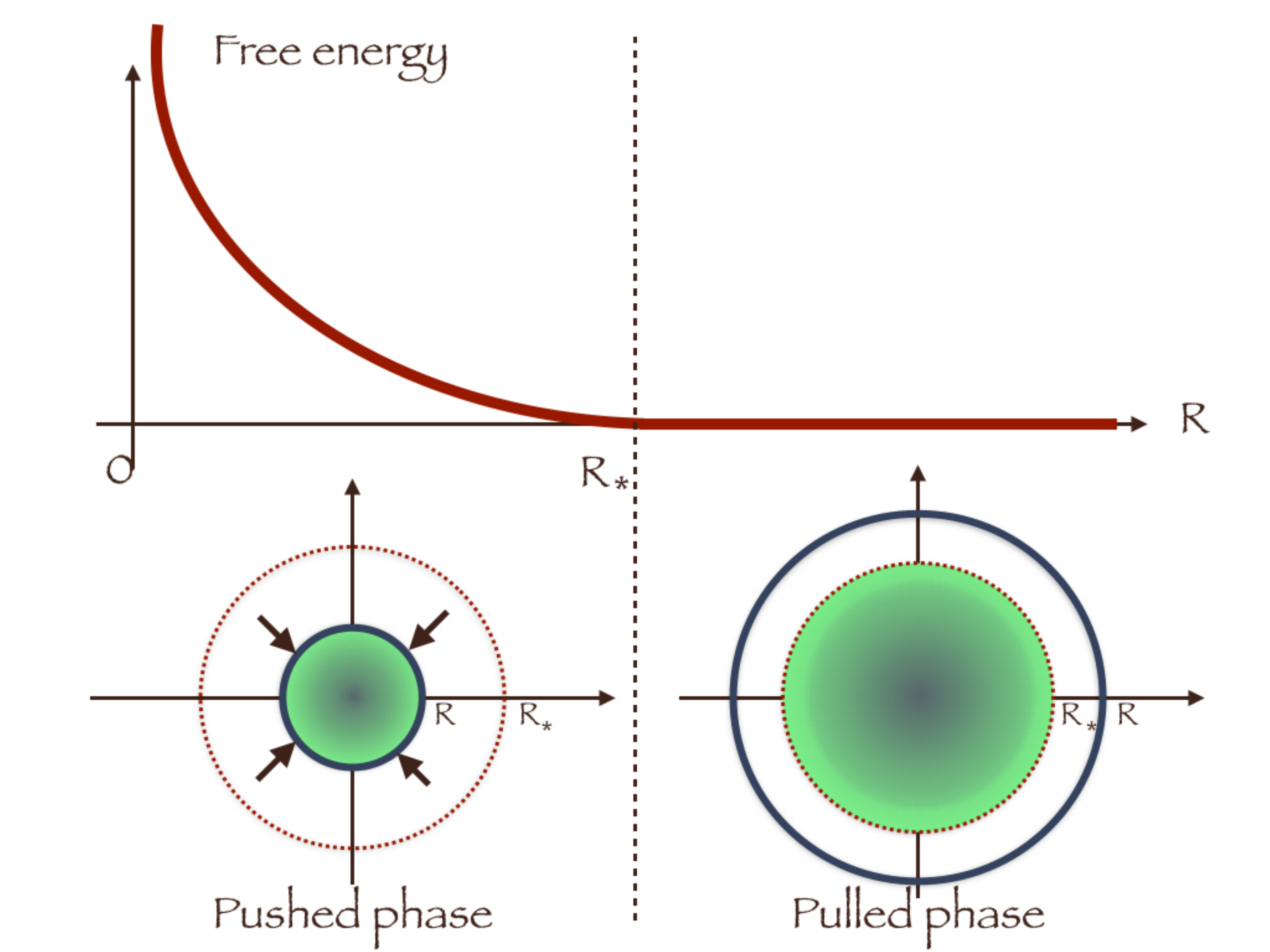}
\caption{
Illustration of the pulled-to-pushed transition for a constrained gas. At equilibrium, the gas is concentrated in a region $\Gamma$ (here the disk within the red dotted line). Right: For $R>R_{\star}$ the constraint (solid blue line) is ineffective, and the free energy is constant. Left: When $R<R_{\star}$, the free energy increases as the gas gets more and more pushed within a much narrower region than it would normally occupy at equilibrium. At the critical point $R=R_{\star}$ the free energy is not analytic.
}
\label{fig:transition}
\end{figure}

Suppose, for instance, that one constrains the system to be completely contained in a certain region, say a ball $B_R\subset\R^n$ of radius $R>0$. If $R$ is large enough so that $\Gamma\subset B_R$,  the constraint is ineffective (\emph{pulled phase}) and the free energy does not change. On the other hand, if $R$ is so small that $\Gamma\not\subset B_R$, then the equilibrium configuration of the system changes (\emph{pushed phase}) and the free energy increases.  It is clear that there exists a critical radius $R_{\star}$ that separates the two phases; for $R> R_{\star}$ the system is in the pulled phase; for $R< R_{\star}$ the system is in a pushed phase. 
In order to go beyond this qualitative picture, sketched in Fig. \ref{fig:transition}, it would be desirable to assign a thermodynamic meaning to this `pulled-to-pushed' phase transition and compute its order.

Examples of this transition abound in the physics literature of particles interacting via logarithmic potential (\emph{log-gas}) in dimension $n=1$ and $n=2$. These systems are related to the eigenvalue statistics of unitarily invariant random matrix models~\cite{Mehta04} (see Section~\ref{sub:log}). 
In those examples, a thermodynamic potential typically turns out to have a discontinuity in the third derivative at the critical value $R_{\star}$. The conclusion is that the `pulled-to-pushed' transition is a \emph{third-order phase transition} (in the sense of Ehrenfest). 

Perhaps, the most popular third-order transitions are the so called Gross-Witten-Wadia~\cite{Gross80,Wadia80} and Douglas-Kazakov~\cite{Douglas93} large-$N$ phase transitions. In both cases, the authors realised that even the simplest integrals involving the unitary group  were non-analytic in the coupling constant. In particular, at the critical value of the coupling constant separating  the weak coupling and the strong coupling phases, the free energy has a jump in the third derivative. These models are naturally mapped onto the statistical physics of log-gases, and the strong-to-weak coupling transition of lattice gauge theories can be translated as a pulled-to-pushed transition for classical particles.
Several other third-order phase transitions have been discovered over the last  decade in many physics problems related at various levels to random matrices, such as the distribution of the
maximum height of non-intersecting Brownian excursions~\cite{FMS11,SMCF}, conductance and shot noise in chaotic cavities~\cite{Cunden15,VMB08,DMTV11}, R\'enyi entanglement entropy of bipartite random pure
states~\cite{Facchi08,NMV10,Facchi10}, wireless telecommunication~\cite{MIMO}, complexity of spin glass landscapes~\cite{FN12}, and discrete log-gases related to random tilings~\cite{Colomo13} (see the review~\cite{Majumdar14} and references therein for more examples and a detailed discussion).

Based on the available evidence so far, the natural question that arises is: to what extent is the third-order pulled-to-pushed phase transition \emph{universal}? In other words, what are the ingredients---dimension, type of interactions, properties of the confining potential, etc.---that are necessary to induce this type of transition? 
In a recent work~\cite{Cunden16s}, three of us suggested that the order of the phase transition should be related to the regularity of the equilibrium measure as a function of the external constraint. In this work, we elaborate further on this idea. The prominent role played by log-gases in $\R$ or $\R^2$ in the literature on the topic might lead to believe 
that the appearance of such weak transitions must be inextricably linked to this specific type of pairwise interaction and/or to low dimensions. Yet, we show here that this is not the case.

More precisely, consider a system of $N$ classical charged particles in $\R^n$. The particles interact via the $d$-dimensional Coulomb potential~$\Phi_d(x)$, the free space Green's function for $-\Delta$ in dimension $d$  (see Eq. (\ref{poisson}) below). Notice that the Coulomb interaction is repulsive and long-range.  To ensure stability in the absence of constraints, the particles are subject to an isotropic confining potential~$V(x)$ which may be generated, e.g., by a fixed neutralising background of opposite total charge $-N$ (this is related to the celebrated \emph{jellium} model of Wigner). 
Here we consider the natural case $n=d$, i.e. \emph{$d$-dimensional Coulomb gases in $\R^d$}. These systems have recently received much attention (see, for instance, the works of Chafa\"i, Gozlan and Zitt~\cite{Chafai14}, Rougerie and Serfaty~\cite{Rougerie15} and Lebl\'e and Serfaty~\cite{Leble15}).

For these Coulomb gas models, we establish that, if the confining  potential $V(x)$ is radially symmetric, a constraint on the volume available to the system induces a \emph{third-order} phase transition, \emph{irrespective of the dimension} and under mild assumptions on $V(x)$. Furthermore, a general formula for the excess free energy in all dimensions is also derived (see Eq. (\ref{eq:rateFgeneral}) below). This third-order transition was established by an explicit computation in a previous work~\cite{Cunden16} for the jellium model in $d=2$ (see also~\cite{Allez14,Jancovici93}). In this paper, we make the most of the isotropy assumption and the basic properties of the $d$-dimensional Coulomb interaction in $\R^d$ to reduce the many-body problem in dimension $d$ to an integrable one-dimensional system.

The paper is organised as follows. In the next Section, we review two examples of third-order phase transitions in random matrix theory. Then in Section~\ref{sec:main}  we define the model and its thermodynamic limit, and we present and discuss the main result. In Section~\ref{sec:proof} we derive the main formula on the large deviation functions.
As an illustrative example, in Section~\ref{sec:jellium} we present in details explicit formulae for the quadratic case, related to random matrices and the jellium model. The final Section is a summary with a pointer to open problems.

\section{Examples of third-order phase transitions for log-gases}
\label{sub:log}

Third-order phase transitions have been observed in one-dimensional and two-dimensional systems with logarithmic repulsion, i.e. for eigenvalues of unitarily invariant matrix models. A rather extensive discussion for Hermitian models is contained in~\cite{Majumdar14}. 
In this section we discuss two paradigmatic examples of random matrices~\cite{Dyson62,Ginibre65,Mehta04}. 
\begin{enumerate}
\item $M$ is a Hermitian matrix of size $N$, whose off-diagonal entries are complex standard Gaussian, independent modulo the symmetry $M_{ij}=\overline{M}_{ji}$, while the diagonal entries are independent real standard Gaussian. The eigenvalues of $M$ are real random variables. This ensemble is named Gaussian Unitary Ensemble (GUE);  
\item  $M$ is a $N\times N$ complex matrix whose $N^2$ entries are independent complex  standard Gaussian. The eigenvalues of $M$ are generically complex. This ensemble is called Ginibre Unitary Ensemble (GinUE). 
\end{enumerate}
The key fact of these ensembles is that the joint probability density of the eigenvalues $(x_1,\dots,x_N)$ of $M$ is  explicitly known~\cite{Mehta04}. After a suitable rescaling on the variances of the entries $M_{ij}$, one finds that
\barr
\mathbb{P}_{N}\left(x_1,\dots,x_N\right)&&=\displaystyle\frac{1}{\mathcal{Z}_{N}}\exp\left({-N\sum_{k=1}^N|x_k|^2}\right)\prod_{i<j}|x_i-x_j|^2\ ,\label{eq:Gaussianjpd}\\
&&\text{where}\quad
\left\{  \begin{array}{l@{\quad}cr} 
x_i\in\R & \mathrm{if} & M\in\text{GUE}\ , \\ 
x_i\in\R^2 & \mathrm{if}  &M\in\text{GinUE}\ . \\ 
\end{array}\right.\nonumber
\earr
(Of course, the value of the normalization constant $\mathcal{Z}_{N}$ is different in the two ensembles. With a slight abuse of notation, we will not distinguish between a measure
and its density, and we identify $\C\simeq\R^2$.)

As remarked by Dyson~\cite{Dyson62}, the joint density~\eref{eq:Gaussianjpd} can be seen as the canonical measure of a system of particles interacting via logarithmic repulsion in a quadratic external potential at  inverse temperature $\beta=2$. In the GUE ensemble, the particles are constrained on the real line ($n=1$); in the GinUE ensemble the particles live on the plane ($n=2$). With this picture in mind, $\mathcal{Z}_{N}$ \emph{in both cases} is the partition function of a $d=2$ Coulomb gas ($-\log|x|$ is the electrostatic potential in dimension \emph{two}):
\be
\mathcal{Z}_{N}=\int_{\R^n}\de x_1\cdots\int_{\R^n}\de x_N\exp\bigg(\sum_{i,j \,:\, i\neq j}\log|x_i-x_j|-N\sum_k |x_k|^2\bigg)\ ,
\ee
where $n=1$ if  $M\in$ GUE, and  $n=2$ if  $M\in$ GinUE. 

Already at this early stage, a moment of reflection brings up a rather curious feature of these particle systems: for the GUE ensemble, there is an evident mismatch between the physical dimension where the particles live ($n=1$) and the Coulomb potential the particles feel ($d=2$). We will come back to this simple observation in the Conclusion section. To avoid confusion, we will always denote as \emph{log-gas} a system of particles in $\R^n$ (for generic $n\geq1$) repelling with a logarithmic pairwise interaction.

For the GUE and GinUE, the eigenvalues empirical distribution $\rho_N=\frac{1}{N}\sum_{k=1}^N\delta_{x_k}$ converges to the celebrated `Wigner law' and `circular law', respectively:
\barr
\rho_{\text{GUE}}(x)&=\frac{1}{\pi}\sqrt{2-x^2}\, 1_{|x|\leq R_{\star}},\quad &R_{\star}=\sqrt{2},\quad(x\in\R)\ ,\label{eq:semicirlce}\\
\rho_{\text{GinUE}}(x)&=\frac{1}{\pi}\, 1_{|x|\leq R_{\star}},\quad &R_{\star}=1,\quad (x\in\R^2)\ .\label{eq:cirlce}
\earr
Here $1_A$ denotes the indicator function of the set $A$, having the value $1$ for all $x\in A$ and $0$ for all $x\notin A$.
The important feature is that, in both cases, the log-gas density concentrates on a bounded region (a ball of radius $R_{\star}$).  It is possible to show that, as $N\to\infty$ (and without any scaling), $\Prob\{\max|x_i|\leq R\}$ converges to a step function: $0$ if $R<R_{\star}$, and $1$ if $R>R_{\star}$. The large-$N$ scaling of these distributions approaching the step function is
\be
\Prob\{\max|x_i|\leq R\}\approx \mathrm{e}^{-N^2F(R)}\ , \label{eq:LDP1}
\ee
where the symbol $\approx$ stands for equivalence at logarithmic scales.
The exponential decay is decorated with a `large deviation function' $F(R)$ which is nothing but the excess free energy of the log-gas constrained to stay in the ball $B_R=\{|x|\leq R\}$:
\barr
F(R)&&=-\lim_{N\to\infty}\frac{1}{N^2}(\log \mathcal{Z}_{N}(R)-\log \mathcal{Z}_{N})\ ,\\
\mathcal{Z}_{N}(R)&&= \int_{B_R}\de x_1 \cdots\int_{B_R} \de x_N\,\mathrm{e}^{\,\sum_{i\neq j}\log|x_i-x_j|-N\sum_k |x_k|^2}\ .
\earr

The computation of the large deviation functions has been carried out explicitly by Dean and Majumdar~\cite{Dean08} for the GUE, and by Cunden, Mezzadri and Vivo~\cite{Cunden16} for the GinUE\footnote{These results are actually valid for any Dyson index $\beta>0$, not just for the GUE and GinUE ($\beta=2$), after a simple rescaling of the eigenvalues. The function $F_{\text{GinUE}}(R)$ is identical to a limiting case of the large deviation function for the index of GinUE computed in~\cite{Allez14}.}.  The final result is
\be
F_{\text{GUE}}(R)=
\left\{  \begin{array}{l@{\quad}cr} 
\displaystyle\frac{1}{16}\left(8R^2-R^4-16\log R-12+8\log2\right) & \mathrm{if} & R\leq R_{\star}\ ,\smallskip \\
0 &  \mathrm{if} & R> R_{\star}\ ,
\end{array}\right.
\label{eq:rateFGUE}
\ee
\be
F_{\text{GinUE}}(R)=
\left\{  \begin{array}{l@{\quad}cr} 
\displaystyle\frac{1}{4}(4R^2-R^4-4\log R-3) & \mathrm{if} & R\leq  R_{\star}\ ,\smallskip \\  
0 &  \mathrm{if} & R> R_{\star}\ .
\end{array}\right.
\label{eq:rateFGinUE}
\ee

The form of the large deviation function is specific to the model ($F_{\text{GUE}}(R)\neq F_{\text{GinUE}}(R)$). 
Nevertheless, in both cases the pulled-to-pushed transition is a third-order phase transition, i.e.,  the excess free energies are non-analytic with a discontinuous third derivative exactly at the critical point $R=R_{\star}$:
\be
\fl\quad F_{\text{GUE}}(R)\sim\frac{\sqrt{2}}{3} (R_{\star}-R)^3 1_{R\leq R_{\star}},
\quad F_{\text{GinUE}}(R)\sim\frac{4}{3}(R_{\star}-R)^3 1_{R \leq R_{\star}},\quad\mbox{as } R\to R_{\star}\ .
\ee 

There is a long list of matrix models exhibiting this third-order singularity. It is therefore tempting to suspect that \emph{non-universal large deviation functions} of generic statistical models share the same \emph{universal critical exponent} in presence of volume constraints. However, it also seems that logarithmic interactions must play a prominent, if not essential, role in this respect. Our findings below will instead show that a logarithmic repulsion is \emph{not} essential to obtaining a third-order phase transition, and the source of this universality must be sought elsewhere.

\section{Large deviations and third-order phase transition for Coulomb gases}
\label{sec:main}

\subsection{Definition of the model}
\label{sub:def}

We consider here a \emph{$d$-dimensional Coulomb gas in $\R^d$}.
The canonical distribution of $N$  charges in dimension $d$ is 
\be
 \mathbb{P}_{d,N}\left(x_1,\dots,x_N\right)=\displaystyle\frac{1}{\mathcal{Z}_{N,\beta}}\mathrm{e}^{-\beta E_d(x_1,\dots,x_N)}\ , 
\label{eq:jpdfC}
\ee
with the energy
\be
E_d\left(x_1,\dots,x_N\right)=\displaystyle\frac{1}{2}\sum_{i,j \,:\,  i\neq j}\Phi_d(x_i-x_j)+N\sum_k V\left(x_k\right),\quad (x_i\in\R^d)\ ,
\label{eq:hamiltonianC}
\ee 
and the partition function
\begin{equation}
\label{eq:partitionfunC}
\mathcal{Z}_{N,\beta}=\!\!\int_{{\R^d}}\!\de x_1\cdots\int_{{\R^d}}\de x_N \mathrm{e}^{-\beta E_d(x_1,\dots,x_N )}\ .
\end{equation}

Here $\Phi_d(x)$  is the Coulomb electrostatic potential in dimension $d$, i.e., the solution of the distributional equation
\be
-\Delta\Phi_d(x)=\Omega_d\delta(x)\qquad (x\in\R^d, \; d\geq1)\ ,\label{poisson}
\ee
where $\Omega_d=2\pi^{d/2} /\Gamma(d/2)$ is the surface area of the unit sphere $S^{d-1}$ in $\R^d$, and  $\Gamma$ is the Euler gamma function. One gets 
\be
\Phi_d(x)=\varphi_d(|x|)\ , \qquad
\varphi_d(r)=
\left\{  \begin{array}{l@{\quad}cr} 
\displaystyle\frac{1}{(d-2)}\frac{1}{r^{d-2}} &  \mathrm{if} & d\neq 2\ ,\\
\\
-\log r &  \mathrm{if} & d=2\ , 
\end{array}\right.
\ee
where $|\cdot|$ denotes the Euclidean norm. To ensure the finiteness of the partition function for sufficiently large $N$, we will require the integrability and the growth conditions 
\be
\int_{\R^d}\exp(-V(x))\de x<\infty\ ,\quad \lim_{|x|\to\infty}\frac{V(x)}{\Phi_d(x)}=+\infty\ .\label{eq:conditions}
\ee 
\subsection{Thermodynamic limit and main result}
\label{sub:thermo}
It is known~\cite{Spohn99} that entropy plays no role at leading order in $N$ in the asymptotics of the partition function. This can be understood easily by recasting~\eref{eq:jpdfC} as
\be
\fl\qquad \mathcal{Z}_{N,\beta}=\!\!\int\limits_{{\R^d}}\!\de x_1\cdots\int\limits_{{\R^d}}\de x_N\exp{\Big(-\beta N\Big[\frac{1}{N}\sum_{i< j}\Phi_d(x_i-x_j)+\sum_k V(x_k)\Big]\Big)}\ ,
\ee
which shows, at least formally, that the limit $N\to\infty$ is a simultaneous thermodynamic  and zero-temperature limit in the mean-field regime. (This readily explains the familiar rescaling in $\beta>0$ of the large deviation functions in random matrix theory.)
The energy of a configuration $(x_1,\dots,x_N)$ can be written in terms of the empirical distribution, 
\be
\rho_N=\frac{1}{N}\sum_{k=1}^N\delta_{x_k}\ ,
\ee 
as
\be
E_d\left(x_1,\dots,x_N\right)=N^2\EE_d[\rho_N]\ ,
\ee  
where the mean field energy functional is defined as
\be
\EE_d[\rho]=\frac{1}{2}\iint_{x\neq y}\Phi_d(x-y)\de\rho(x)\de\rho(y)+\int V(x)\de\rho(x)\ .
\label{eq:energyfunct}
\ee
Under the  assumptions~(\ref{eq:conditions}) on $V(x)$, it is known (see, for instance~\cite{Chafai14}) that the functional $\EE_d$ has weak-$\ast$ compact level sets (and is thus lower semicontinuous) and that $\EE_d$  is strictly convex where it is finite. The proof of the convexity is quite standard in $d=1$ and $d=2$. For $d\geq3$, the proof is based on the fact that the external potential part is linear while the interaction potential can be written as conic combination of Gaussian kernels
\be
\varphi_d(r)=
\frac{\Omega_d}{4\pi^{d/2}}\int_{0}^{\infty}t^{\frac{d-4}{2}}\mathrm{e}^{-r^2t}\de t\ .
\ee 
Compactness and strict convexity imply that $\EE_d$ has a unique minimiser $\rho_{\text{eq}}$, called \emph{equilibrium measure},  among the probability measures on $\R^d$.  It turns out that, in the large-$N$ limit, the empirical distribution $\rho_N$  converges to the deterministic equilibrium measure $\rho_{\text{eq}}$~\cite{Chafai14,Leble15,Spohn99,Rougerie15}. 

In the following, we consider the case of a radial external potential $V(x)=v(|x|)$ satisfying the following hypotheses.

\smallskip
\paragraph{Assumptions A-1:} The external radial potential $v(r)$ ($r\geq0$) is of class $C^3$. Moreover, we assume that $v(r)$ and  $r^{d-1}v'(r)$ are both strictly increasing (the last is true, e.g., if $v(r)$ is strictly convex). 
\smallskip

 Note that the problem is radially symmetric and this symmetry is inherited by~$\rho_{\text{eq}}$. By our assumptions on $v(r)$, it follows that  $\rho_{\text{eq}}$ is supported on a ball. Indeed,  introducing spherical coordinates $x= r \omega \in\R^d$, with $r\geq0$ and $\omega\in S^{d-1}$, 
 by an application of Gauss's law,
\begin{equation}
\de\rho_{\text{eq}}(x)=\frac{1}{\Omega_d}\Delta V(x)\de x\ ,
\end{equation} 
we get
\be
\fl \qquad \de x=r^{d-1}\de r \de \omega,\qquad \Delta V(x)=\frac{\de^2v(r)}{\de r^2}+\frac{d-1}{r}\frac{\de v(r)}{\de r}=\frac{1}{r^{d-1}}(r^{d-1} v'(r))'\ ,
\ee
and thus
\be
\de\rho_{\text{eq}}
=\frac{1}{\Omega_d}(r^{d-1} v'(r))' \,1_{r\leq R_{\star}}\,\de r\de \omega\ .
\label{eq:saddledensity}
\ee
The equilibrium measure is concentrated on a ball of radius $R_{\star}$, which is given by the unique positive solution of 
\be
v'(R_{\star})R_{\star}^{d-1}=1\ ,
\label{eq:Rstar}
\ee    
a direct consequence of normalization $\int \de \rho_{\text{eq}} = 1$.

Following the same line of reasoning of Section~\ref{sub:log}, the probability that the Coulomb gas is contained in a ball of radius $R$ converges to a step function 
\be
\Prob\{\max|x_i|\leq R\}\stackrel{N\to\infty}{\longrightarrow}
\left\{  \begin{array}{l@{\quad}cr} 
0& \mathrm{if} & R< R_{\star}\ , \\  
1 &  \mathrm{if} & R\geq R_{\star}\ .
\end{array}\right.
\label{eq:limit}
\ee 
On the other hand, similarly to~\eref{eq:LDP1}, one expects
\be
\Prob\{\max|x_i|\leq R\}\approx \mathrm{e}^{-\beta N^2F_d(R)}\ ,\label{eq:LDP}
\ee
where the large deviation function $F_d(R)$ decorating the asymptotics~\eref{eq:LDP} is the excess free energy of the constrained $d$-dimensional Coulomb gas
\be
F_d(R)=
-\lim_{N\to\infty}\frac{1}{\beta N^2}\Big(\log \mathcal{Z}_{N,\beta}(R) - \mathcal{Z}_{N,\beta} \Big)\ ,
\label{eq:freeen}
\ee
where $\mathcal{Z}_{N,\beta}$ is the free partition function~(\ref{eq:partitionfunC}), while 
\begin{equation}
\label{eq:partitionfunCR}
\mathcal{Z}_{N,\beta}(R)= \int_{B_R}  \de x_1 \cdots \int_{B_R}  \de x_N \,\mathrm{e}^{-\beta E_d(x_1,\dots,x_N)}
\end{equation}
is the partition function of the Coulomb gas constrained to stay inside the ball of radius $R$ 
\begin{equation}
\label{eq:BRdef}
B_R=\{x\in \mathbb{R}^d \,:\, |x|\leq R\}\ .
\end{equation}
From~\eref{eq:limit} and~\eref{eq:LDP}, we infer that 
$F_d(R)$ is identically zero for $R\geq R_{\star}$ and strictly positive when $R<R_{\star}$. Hence, the excess free energy is non-analytic at the critical value $R=R_{\star}$ that separates the pulled and the pushed phases. What is the order of this transition?
Our main result is the existence of a third-order phase transition with respect to the parameter $R$. The precise statement is as follows.
\medskip

\noindent\fbox{\begin{minipage}{1\columnwidth}
\smallskip
\paragraph{Main result.}

Let the radial potential $V(x)=v(|x|)$ satisfy the coercivity conditions~(\ref{eq:conditions}) and the Assumptions~A-1. Then, 
\begin{itemize}
\item[(i)] the excess free energy~(\ref{eq:freeen}) is given by the explicit formula: 
\be
F_d(R)=\frac{1}{2} \!\!\! \int\limits_{\min\{R,R_{\star}\}}^{R_{\star}} \left(r^{d-1}v'(r)^2-2v'(r) -\varphi_d'(r)\right)\,\de r\ ,
\label{eq:rateFgeneral}
\ee
\item[(ii)] the system undergoes a third-order phase transition at $R=R_{\star}$, i.e., \mbox{$F_d(R)\in C^2(0,\infty)$}, and at the critical value the third derivative is discontinuous:
\barr
F_d(R_{\star})=0,\quad F'_d(R_{\star})=0, \quad F''_d(R_{\star})&&=0\ ,\\
\text{and}\,\,\,  \lim_{R \uparrow R_{\star}}F'''_d(R)< \lim_{R\downarrow R_{\star}}F'''_d(R)&&=0\ .
\earr
\end{itemize}
\end{minipage}}
\smallskip

If we accept the claim (i), we can prove the third-order phase transition (ii) as follows. $F_d(R)$ given in~\eref{eq:rateFgeneral} is manifestly continuous. 
Note that $F_d(R)$ and all its derivatives $F_d^{(k)}(R)$ are identically zero for $R> R_{\star}$. It remains to study the behaviour of $F_d^{(k)}(R)$ as $R\uparrow R_{\star}$. 
From~\eref{eq:rateFgeneral} we have that, for $R<R_{\star}$,
\barr
F_d'(R)&=&\frac{\varphi_d'(R)}{2}+v'(R)-\frac{1}{2}R^{d-1}v'(R)^2\ ,\label{eq:1st}\\
F_d''(R)&=&\frac{\varphi_d''(R)}{2}+v''(R)-\frac{d-1}{2}R^{d-2}v'(R)^2-R^{d-1}v'(R)v''(R)\ ,\label{eq:2nd}\\
F_d'''(R)&=&\frac{\varphi_d'''(R)}{2}+v'''(R)-\frac{(d-1)(d-2)}{2}R^{d-3}v'(R)^2\nonumber\\
&-&2(d-1)R^{d-2}v'(R)v''(R)-R^{d-1}v''(R)^2-R^{d-1}v'(R)v'''(R)\ .\label{eq:3rd}
\earr
 We also need the first three derivatives of the Coulomb potential
\be
\varphi_d'(r)=-\frac{1}{r^{d-1}},\quad\varphi_d''(r)=-\frac{d-1}{r^{d}},\quad\varphi_d'''(r)=\frac{d(1-d)}{r^{d+1}}\ .
\label{eq:varphider}
\ee
Therefore, from~\eref{eq:1st}, using~\eref{eq:Rstar} and \eref{eq:varphider} we find
\barr
\lim_{R\uparrow R_{\star}}F'_d(R)&=&\frac{1}{2}\left(\varphi_d'(R_{\star})+v'(R_{\star})\right)=
\frac{1}{2}\left(-\frac{1}{R_{\star}^{d-1}} +v'(R_{\star})\right)=0\ .
\label{eq:balance_edge}
\earr 
For the second derivative, using again~\eref{eq:Rstar} and~\eref{eq:varphider}, we get
\barr
\lim_{R\uparrow R_{\star}}F''_d(R)&=&\frac{\varphi_d''(R_{\star})}{2}-\frac{d-1}{2R}v'(R_{\star})=\frac{d-1}{2}\left(\frac{1}{R_{\star}^d}-\frac{v'(R_{\star})}{R_{\star}}\right)=0\ .
\earr
We analyse now the third derivative. Proceeding as before we find
\be
\fl\qquad \lim_{R\uparrow R_{\star}}F'''_d(R)=\frac{\varphi_d'''(R_{\star})}{2}-\frac{(d-1)(d-2)}{2R_{\star}^{d+1}}-2\frac{(d-1)}{R_{\star}^{d+1}}v''(R_{\star})-\frac{v''(R_{\star})^2}{R_{\star}^{1-d}}\ .
\label{eq:thirdstar}
\ee
If we write explicitly the condition that $r^{d-1}v'(r)$ is strictly increasing we have
\be
v''(r)>\frac{1-d}{r}v'(r),\quad r>0\ .
\ee
Hence we get $v''(R_{\star})>(1-d)/{R_{\star}^{d}}$ and using~\eref{eq:varphider} we conclude that
\be
\fl \qquad \lim_{R\uparrow R_{\star}}F'''_d(R) < \frac{d(1-d)}{2R_{\star}^{d+1}}-\frac{(d-1)(d-2)}{2R_{\star}^{d+1}}-2\frac{(d-1)^2}{R_{\star}^{d+1}}+\frac{(d-1)^2}{R_{\star}^{d+1}} = 0\ .
\ee
This concludes the proof of the third-order transition
\be
\boxed{
F_d(R)\simeq  (R_{\star}-R)^31_{R\leq R_{\star}},\quad\text{as } R\to  R_{\star}\ .
\label{eq:exponent}
}
\ee 
It remains to prove the claim (i) on the explicit form of the large deviation function. 
\paragraph{Remarks.} The assumptions on the potential $v(r)$ are not technical. Of course, we require that $v\in C^3$ in order to make sense of the derivatives of $F_d(R)$ for $R<R_{\star}$ [see~\eref{eq:1st}-\eref{eq:3rd}]. The condition that $v(r)$ and $r^{d-1}v'(r)$ are strictly increasing implies that the support of equilibrium measure (pulled phase) is simply connected (a ball). It is known that the `transitions' when the pulled phase is supported on a non-simply connected region display, in general, different exponents (see, e.g., multi-cut Hermitian matrix models).

The exponent 3 in~\eref{eq:exponent} is therefore `universal'. [The non-universal constant in front of~\eref{eq:exponent} is the right hand-side of~\eref{eq:thirdstar}.]   For $d=2$, the explicit computation of $F_d(R)$ in the case of quadratic potential (Ginibre ensemble in random matrix theory) has been done in~\cite{Cunden16}. In that paper, the authors posed a question on the universality of the third-order transition for normal matrix models. The answer is `Yes'.
\section{Free energy for radial potentials}
\label{sec:proof}
In order to compute the large deviation function~\eref{eq:freeen}, one needs to extract the leading order in the asymptotics of the partition function of the constrained Coulomb gas~(\ref{eq:partitionfunCR}).
Recently, the large-$N$ asymptotics of the free energies of Coulomb gases in dimension $d$ with quite generic potential has been investigated in great detail~\cite{Chafai14,Leble15,Spohn99,Rougerie15,Weigmann06}. It has been proved that, for large $N$, the free energy is dominated by the minimum of the energy functional $\EE_d$ in~\eref{eq:energyfunct}. Then, the following saddle-point approximation holds true
\be
-\lim_{N\to\infty}\frac{1}{\beta N^2}\log \mathcal{Z}_{N,\beta}(R)= \EE_d[\rho_R]\ ,\label{eq:saddle}
\ee
where $\rho_R$ is the equilibrium measure of the constrained Coulomb gas, i.e., the minimiser of the mean field functional over the set of probability measures supported on the ball of radius $R$
\be
\EE_d[\rho_R]=\min\bigg\{\EE_d[\rho]\; \colon\;  \rho\geq0, \; \int_{B_R} 
\hspace{-2mm}\de\rho(x)=1\bigg\}\ .
\ee
Therefore, the excess free energy~(\ref{eq:freeen}) is given by
\be
F_d(R)=-\lim_{N\to\infty}\frac{1}{\beta N^2}\log \frac{\mathcal{Z}_{N,\beta}(R)}{\mathcal{Z}_{N,\beta}}=\EE_d[\rho_R]-\EE_d[\rho_{\text{eq}}]\ .
\ee
Thus the technical problem is i) to find the minimiser $\rho_R$, and ii) compute the minimum energy $\EE_d[\rho_R]$.  
\subsection{Finding the equilibrium measure}
To solve the problem, we exploit the radial symmetry of the system and the fundamental properties of the Coulomb potential. First, we define the total energy density  of the constrained gas at $x\in\R^d$ 
\be
\varepsilon_d(x,R)=\int\Phi_d(x-y)\, \de\rho_R(y)+V(x)\ .
\label{eq:endensity}
\ee
Observe that $\varepsilon_d(x,R)$ is the electrostatic potential at $x$ jointly
created by the distribution of charge $\rho_R$ and the external potential $V$.
The unknown measure $\rho_R$ satisfies the condition of electrostatic equilibrium~\cite{Chafai14}
\be
\left\{  \begin{array}{l@{\quad}cr} 
\varepsilon_d(x,R)\geq	 C_R&&\text{for $|x|\leq R$}\ ,  \\ 
\varepsilon_d(x,R)=C_R&&\text{for $x$ in the support of $\rho_R$}\ ,
\end{array}\right.
\label{eq:electrequi}
\ee
for some constant $C_R$ (independent of $x$).  The `Euler-Lagrange equations' \eref{eq:electrequi} can be understood, in electrostatic terms, as follows. The fact that $\varepsilon_d(x,R)$ is constant
inside the support of $\rho_R$ indicates that the electric field ($-\nabla\varepsilon_d(x,R)$) is zero inside this support. The fact
that $\varepsilon_d(x,R)$ takes greater values outside this support implies that the charges are confined within
the support, and that taking any small amount of charge outside the support would increase
the total electrostatic energy of the system. 
In the unconstrained problem, the equilibrium measure $\rho_{\text{eq}}$ is supported on the ball of radius $R_{\star}$ [see~\eref{eq:saddledensity}-\eref{eq:Rstar}].
Hence, as long as $R\geq R_{\star}$, the constraint is ineffective and the functional is minimised by $\rho_R=\rho_{\text{eq}}$. 

Let us consider now the case $R<R_{\star}$, when the Coulomb gas gets pushed by the volume constraint. The minimiser can be identified by imposing two basic conditions valid at electrostatic equilibrium:  the Gauss's law must be satisfied and any excess of charge gets localised at the surface of a conductor.
Hence, the volume constraint does not change the charge distribution in the interior of the ball and we have
\be
\rho_R(x)=\rho_{\text{eq}}(x)\quad\text{for  $|x|<R$}\ .
\ee
 Of course $\int_{|x|<R}\de\rho_R(x)=\int_{|x|<R}\de\rho_{\text{eq}}(x)<1$. The missing charge is localized on the surface and, by symmetry, is uniformly distributed on the sphere of radius $R$. Using spherical coordinates, the radial and angular distributions factorise and eventually we find 
\be
\fl\qquad \de\rho_{R}(r,\omega)=\frac{1}{\Omega_d} \left[(r^{d-1} v'(r))'\, 1_{r\leq R}
+\left(1-R^{d-1} v'(R)\right)\delta(r-R) \right]\de r \de \omega\ ,
\label{eq:equilibriummeas}
\ee
valid when $R<R_{\star}$. In the pushed phase, the Coulomb gas equilibrium measure has a singular component (non-fluid phase). At $R=R_{\star}$, using~\eref{eq:Rstar} one recovers $\rho_{\text{eq}}$. 

The correctness of~\eref{eq:equilibriummeas} can be ascertained by checking the equilibrium conditions~\eref{eq:electrequi}.
Here we need the following electrostatic integration formula
\be
\frac{1}{\Omega_d}\int_{S^{d-1}}\varphi_d(|x- r \omega|)\, \de \omega =\,\varphi_d(\max \{|x|, r\})\ .
\label{eq:intformula}
\ee
(The identity is a one-line proof for $d=1$, a classical formula in complex analysis for $d=2$ and a familiar fact for electrostatics in dimension $d=3$. In fact, it is possible to prove~\eref{eq:intformula} for generic $d\geq1$ by using the same textbook argument valid in $\R^3$ based on Gauss's law. See~\cite[Thm.~9.7]{LiebLoss}.)

Using~\eref{eq:intformula}, we easily verify that~\eref{eq:equilibriummeas} satisfies the condition of constant energy density $\varepsilon_d(x,R)$ on the support of $\rho_R$. For $|x|\leq R$, we have
\barr
\varepsilon_d(x,R)&=& v(|x|)+\int_0^{\infty}\varphi_d(\max (|x|, r))(r^{d-1} v'(r))'\, 1_{r\leq R}\,\de r\nonumber \\
&&\qquad +\left(1-R^{d-1} v'(R)\right)\int_0^{\infty}\varphi_d(\max (|x|, r))\delta(r-R)\,\de r\nonumber \\
&=&v(|x|)+\varphi_d(|x|)\int_0^{|x|}(r^{d-1} v'(r))'\,\de r+\int_{|x|}^{R}\hspace{-1mm}\varphi_d(r)(r^{d-1} v'(r))'\,\de r\nonumber\\
&&\qquad +\left(1-R^{d-1} v'(R)\right)\varphi_d(R)\nonumber\\
&=&\varphi_d(R)+v(R)\ ,
\label{eq:energy5}
\earr
where we applied (in order) the electrostatic formula~\eref{eq:intformula}, and integration by parts.
\subsection{Computation of the large deviation function}
The next step is to compute the minimum energy $\EE_d[\rho_R]$. The simplest way is by observing that 
\be
\EE_d[\rho_R]=\frac{1}{2}\int_{B_R}\bigg(\varepsilon_d(x,R)+V(x)\bigg)\de\rho_R(x)\ .
\ee
Since, from~\eref{eq:energy5}, $\varepsilon_d(x,R)=\varphi_d(R)+v(R)$ for $|x|\leq R$, and $\int_{B_R}\de\rho_R(x)=1$, using~\eref{eq:equilibriummeas} we  find that, for $R<R_{\star}$,
\barr
\EE_d[\rho_R]
=\frac{1}{2}\varphi_d(R)+v(R)-\frac{1}{2}\int_{0}^Rr^{d-1}v'(r)^2\de r\ .
\label{eq:energyfinal}
\earr
Thus the large deviation function is given by 
\barr
F_d(R)&=&\EE_d[\rho_R]-\EE_d[\rho_{\text{eq}}] = \EE_d[\rho_R]-\EE_d[\rho_{R_\star}] 
\nonumber\\
&=& \int_{R_\star}^R \left[\frac{1}{2} \varphi_d'(r)+v'(r)-\frac{1}{2}r^{d-1}v'(r)^2\right] \de r\ .
\earr 
Note again that $\rho_R=\rho_{\text{eq}}$ for $R\geq R_{\star}$,  and hence $F_d(R)=0$. Therefore $F_d(R)$ can be written as~\eref{eq:rateFgeneral} as claimed. This concludes the proof of part (i) of the main result.

\section{Quadratic confinement: random matrices and the jellium model}
\label{sec:jellium}
\begin{figure}[t]
\centering
\includegraphics[width=.6\columnwidth]{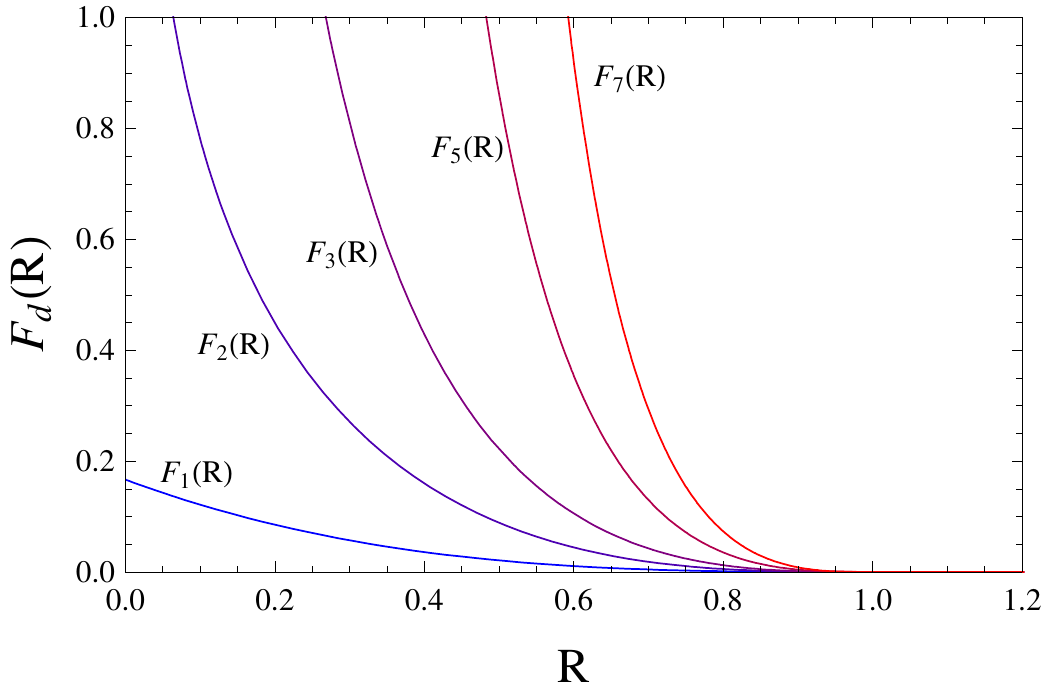}
\caption{Excess free energy $F_d(R)$ (see Eq.~\eref{eq:rateFquad}) for a $d$-dimensional Coulomb gas in $\R^d$ in a quadratic potential, for various values of $d\geq1$. Note that $F_d(R)=0$ for $R$ greater than the critical radius (here $R_{\star}=1$).
}
\label{fig:1}
\end{figure}
In this section, we specialise the general result to the case of a quadratic external potential 
\be
V(x)=\frac{1}{2}|x|^2\ . 
\ee
As already discussed in Section~\ref{sub:log}, in dimension $d=2$, the log-gas in a quadratic potential is statistically equivalent to the eigenvalues of the complex Ginibre ensemble (GinUE). In fact, the Coulomb gas in a quadratic potential is also related to the \emph{jellium} and has a long history in the tradition of classical and quantum statistical physics in dimension $d=1,2$ and $3$ (see~\cite{Alastuey81,Baxter62,Forrester98,Jancovici93,Kunz74,Lieb75,Wigner34}).  The jellium model was invented by Wigner~\cite{Wigner34} to investigate qualitative properties of electrons in metals and plasmas. Wigner's approximation is quite simple:  instead of studying electrons hosted in an ion lattice,  one considers an electron gas embedded in a positive continuum and uniform gel (a `positive jelly').  It is a simple exercise to show that a uniformly charged ball generates a quadratic electric potential inside the ball.

For this model in generic dimension $d$, it is easy to see that, for large $N$, the normalised charge density at equilibrium converges to the uniform measure on the unit ball (the critical radius is $R_{\star}=1$)
\be
\rho_{\text{eq}}(x)\de x=\frac{d}{\Omega_d}1_{|x|\leq1}\de x\ .
\ee 
This is the unique distribution that neutralises the uniform background, and is the $d$-dimensional counterpart of the circular law of random matrix theory ($d=2$). 
In presence of a volume constraint, the minimiser of the energy $\EE_d$ among the probability measures supported on the ball of radius $R$
is given by~\eref{eq:equilibriummeas} with $v(r)=r^2/2$. Explicitly we have
\be
\fl\qquad \de\rho_R(r,\omega)=
\left\{  \begin{array}{l@{\quad}cr} 
\displaystyle\frac{1}{\Omega_d}\left[d\, r^{d-1}1_{r\leq R} +(1-R^d) \delta(r-R)\right]\de r\de\omega & \mathrm{if} & R\leq1\ ,\medskip\\  
\displaystyle\frac{1}{\Omega_d} d \, r^{d-1}1_{r\leq 1}\, \de r\de \omega &  \mathrm{if} & R>1\ .
\end{array}\right.
\label{eq:saddledens}
\ee

Using the general result \eref{eq:rateFgeneral}, we eventually find the unified formula for the large deviation function in generic dimension (see Fig. \ref{fig:1})
\be
\fl \qquad F_d(R)=
\left\{  \begin{array}{l@{\quad}cr} 
\displaystyle\frac{R^{2-d}}{2(d-2)}-\frac{R^{2+d}}{2(d+2)}+\frac{R^2(d-2)(d+2)-d^2}{2(d-2)(d+2)} & \mathrm{if} & R\leq1\ , \medskip\\  
0 &  \mathrm{if} & R>1\ .
\end{array}\right.
\label{eq:rateFquad}
\ee
(for $d=2$ we take $F_2(R)=\lim_{d\to2}F_d(R)$).

For instance, we have
\barr
F_1(R)&&=\frac{1}{6}(1-R)^3\ ,\label{eq:F1}\\
F_2(R)&&=\frac{1}{8}(4R^2-R^4-\log R^4-3)\ ,\label{eq:F2}\\
F_3(R)&&=\frac{1}{2R}-\frac{1}{5}R^5+\frac{1}{2}R^2-\frac{9}{5}\ ,\label{eq:F3}
\earr
for $R\leq1$. For $d=2$, Eq.~\eref{eq:F2} agrees with the result in~\cite{Cunden16}. 
For all $d\geq1$:
\be
F_d(R)\sim\frac{d^2}{6}(1-R)^3\,1_{R\leq1}\quad \text{as $R\to 1$}\ .
\ee

\section{Conclusions}
\label{sec:concl}

From a wide perspective, Coulomb gases  are, perhaps, the simplest example of statistical mechanical systems for which it is possible to assign a thermodynamic meaning to and to characterise in full the `pulled-to-pushed' transition.  The key elements in the proof of our result are the basic properties of the Coulomb interaction and the isotropy of the models considered. As a consequence, the equilibrium measure of those systems can be written rather explicitly, even in presence of hard constraints. The difficulty in trying to generalise our findings to non-isotropic Coulomb gases is that explicit formulae for the equilibrium measure are not readily available. 

We conclude with some open problems and a couple of further remarks, inspired by our discussion in Section~\ref{sub:log}:
\begin{enumerate}
\item The excess free energy of the pushed phase provides the \emph{left} large deviation tail of $\Pr\{\max|x_i|\leq R\}$, when $R\leq R_{\star}$. A standard argument~\cite{Cunden16,Johansson99,Majumdar09,Nadal11} allows to obtain the \emph{right} large deviation tail as  energy cost
in pulling one particle from its equilibrium position inside the support of the equilibrium measure and relocating it in a generic position at distance $R>R_{\star}$ (`pulled' phase). Skipping the details of the computations, one finds
\barr
\Prob\{\max|x_i|\leq R\}\approx \mathrm{e}^{-N^2F_d(R)}\quad&\text{if}&\quad R\leq R_{\star}\ ,\\
\Prob\{\max|x_i|\geq R\}\approx \mathrm{e}^{-N H_d(R)}\quad&\text{if}&\quad R\geq R_{\star}\ ,
\label{eq:finalrateF}
\earr
where $F_d(R)$ is given in~\eref{eq:rateFgeneral}, and 
\be
H_d(R)=\!\!\int\limits_{R_{\star}}^{\max\{R,R_{\star}\}}\!\!\!(\varphi_d'(r)+v'(r))\,\de r\ .
\ee
While the functions $F_d(R)$ and $H_d(R)$ govern the rate of \emph{atypical} fluctuations away from the unconstrained situation, the question of \emph{typical} fluctuations is also interesting. One indeed expects the existence of scaling constants $a_N$, $b_N$ (possibly dependent on $d$) and a cumulative function $G_d(t)$ (the latter independent of $N$) such that 
\be
\Prob\{a_N+b_N\max|x_i|\leq t\}\to G_d(t),\quad \text{as $N\to\infty$}\ .
\ee
In other words, the scaling function $G_d(t)$ describes the typical fluctuations of $\max|x_i|$ in a neighbourhood of the critical point $R_{\star}$. In a certain sense, $G_d$ is the crossover function between the large deviation functions $F_d$ and $H_d$ that describe fluctuations of order $\mathcal{O}(1)$. 

It would be interesting to compute $G_d(t)$ explicitly for various $d$ and address the question of its (microscopic) universality.
For $d=2$ (log-gas in the plane), it is known that $G_{d=2}(t)$ is the Gumbel distribution~\cite{Chafai14b,Rider03} (again under the Assumptions~A-1 on $V(x)$). To the best of our knowledge, the scaling function $G_d(t)$  for $d\neq2$ has not
been investigated. [Note that for the log-gas on the line (e.g. the GUE) the scaling function is a squared Tracy-Widom \cite{Dean16,Edelman15}.]
\item In the GUE model, the equilibrium density (the semicircle law) is continuous and vanishes as a square root $(R_{\star}^2-x^2)^{1/2}$ at the edges~\eref{eq:semicirlce} as long as $R>R_{\star}$; in the `pushed' phase $R<R_{\star}$, the density diverges as $(R^2-x^2)^{-1/2}$ at the pushing walls $\pm R$. For the GinUE ensemble, the equilibrium density (the circular law) is uniform in the unit circle~\eref{eq:cirlce}, and acquires a singular component at the pushing constraining walls. Clearly, the phenomenology of constrained densities for log-gases in dimension $n=1$ and $n=2$ is quite different (see~\cite{Cunden16,Dean08} for details). Nevertheless the critical exponent is the same.

Can a general relation be found between the behaviour of the equilibrium density at the edges and the critical exponent?
An intriguing `scaling argument' was put forward to justify the critical exponent for eigenvalues of Hermitian random matrices (see the review by Majumdar and Schehr~\cite{Majumdar14}). The argument, based on a matching between `typical' and `large' deviations, predicts that if  the equilibrium density (without constraints) vanishes as  $(R_{\star}-x)^{\gamma}$ at the edge, then the excess free energy at the critical point behaves as $F(R)\simeq (R_{\star}-R)^{2(1+\gamma)}$. For off-critical Hermitian random matrices, when $\gamma=1/2$, one indeed recovers the exponent $3$. Unfortunately, this argument seems unable to explain the occurrence of a third-order phase transition in either the GinUE ensemble, or the $d$-dimensional Coulomb gases considered in this paper  (where the equilibrium density is discontinuous at the edge). Hence, the quest for a more viable connection between these seemingly unrelated objects is still open.

\item We noticed in Section \ref{sub:log} that for unitary invariant random matrices a mismatch occurs between the dimension of the Coulomb potential ($d=2$) and the physical dimension the system lives in ($n=1$). Yet, the critical exponent is still 3 (although this has only been proved for a quadratic confining potential). It would be interesting to understand if this feature persists in higher dimensions: for a gas in $\R^n$ interacting through the $d$-dimensional Coulomb repulsion (the solution $\Phi_d$ of the $d$-dimensional distributional equation (\ref{poisson})), is the critical exponent always 3? One may imagine to fill in the following table
\begin{center}
\begin{tabular}{c|ccccc}
&\multicolumn{5}{c}{{\bf Coulomb Interaction} $\Phi_d$}\\
{\bf Physical space} $\R^n$ & 1 & 2 & 3 & 4 & $\hspace{-7mm}\cdots$ \\\hline
1 & $\checkmark$ & $\underline{\checkmark}$ &  &  \\   
2 &  & $\underline{\checkmark}$ &  & \\   
3 &  &  & $\checkmark$ &  \\   
4 &  &  &  & $\checkmark$ \\   
$\vdots$ & & & & &$\hspace{-7mm}\ddots$\\
\end{tabular}
\end{center}
The underlined checkmarks correspond to the log-gases living in dimension $n=1$ and $n=2$  (e.g. the GUE and GinUE, respectively).
In the present work, we have covered the diagonal situation $n=d$ (checkmarks).
The above discussion suggests that for $n<d$ the singularity of the pushed density at the constraining walls should become milder and milder as one moves away from the diagonal $n=d$, but whether the transition is still third-order remains an interesting open problem. 

\end{enumerate}

More generally, the main moral to be drawn from the above is that the analytic properties of the excess free energy of constrained particle systems with repulsive interaction may be universal, i.e., to some extent independent of the detailed behaviour of the constrained equilibrium density. 
One could investigate this point by considering particle systems with interaction kernel $G(x)$ which is the fundamental solution $D G(x)=\delta(x)$ of some differential operator $D$ (in this paper $D=(-1/\Omega_d)\Delta$ is the  Laplacian). This however is another story which will have to await a future work. 
 It may not be too much to hope that the method and the ideas developed here will find useful applications in other statistical models.

\ack

Research of FDC is supported by EPSRC Grant  EP/L010305/1 and by ERC Advanced Grant 669306. ML acknowledges support by Cohesion and Development Fund 2007-2013 - APQ Research Puglia Region ``Regional program supporting smart specialization and social and environmental sustainability - FutureInResearch''. PF acknowledges support by Istituto Nazionale di Fisica Nucleare (INFN) through the project ``QUANTUM''. FDC, PF and ML acknowledge support from the Italian National Group of Mathematical Physics (GNFM-INdAM). 
PV acknowledges the stimulating research  environment provided by the EPSRC Centre for Doctoral Training in Cross-Disciplinary Approaches to Non-Equilibrium Systems (CANES, EP/L015854/1).

\vspace{1cm}


\begin{thebibliography}{99}
\bibitem{Alastuey81} A. Alastuey and B. Jancovici,
\emph{On the classical two-dimensional one-component Coulomb plasma},
J. Physique {\bf 42}, 1-12 (1981).

\bibitem{Allez14} R. Allez, J. Touboul and G. Wainrib,
\emph{Index distribution of the Ginibre ensemble}, 
J. Phys. A: Math. Theor. {\bf 47}, 042001 (2014).


\bibitem{Baxter62} R. J. Baxter,
\emph{Statistical mechanics of a one-dimensional Coulomb system with
a uniform charge background},
Proc. Camb. Phil. Soc. {\bf 59}, 779 (1963).


\bibitem{Chafai14}
D. Chafa\"{i}, N. Gozlan and P.-A. Zitt,
\emph{First-order global asymptotics for confined particles with singular pair repulsion},
Ann. Appl. Probab. {\bf 24}, 2371-2413 (2014).


\bibitem{Chafai14b}
D. Chafa\"{i} and S. P\'ech\'e,
\emph{A Note on the Second Order universality at the Edge of Coulomb Gases on the Plane},
J. Stat. Phys. {\bf 156}, 368-383 (2014).


\bibitem{Colomo13}
F. Colomo and A.~G. Pronko,
\emph{Third-order phase transition in random tilings},
Phys. Rev. E {\bf 88}, 042125 (2013).

\bibitem{Cunden15}
F. D. Cunden, P. Facchi and P. Vivo,
\emph{Joint statistics of quantum transport in chaotic cavities},
 EPL {\bf 110}, 50002  (2015).

\bibitem{Cunden16s}
F. D. Cunden, P. Facchi and P. Vivo,
\emph{A shortcut through the Coulomb gas method for spectral linear statistics on random matrices},
J. Phys. A: Math. Theor. {\bf 49}, 135202  (2016).

\bibitem{Cunden16}
F. D. Cunden, F. Mezzadri and P. Vivo,
\emph{Large Deviations of Radial Statistics in the
Two-Dimensional One-Component Plasma},
J. Stat. Phys. {\bf 164}, 1062-1081 (2016).

\bibitem{DMTV11}
K. Damle, S. N. Majumdar, V. Tripathi and P. Vivo, 
\emph{Phase Transitions in the Distribution of the Andreev Conductance of Superconductor-Metal Junctions with Multiple Transverse Modes},
Phys. Rev. Lett. {\bf 107}, 177206 (2011). 

\bibitem{Dean08} D. S. Dean and S. N. Majumdar,
\emph{Large Deviations of Extreme Eigenvalues of Random Matrices}, 
Phys. Rev. Lett. {\bf 97}, 160201 (2006);  
\emph{Extreme value statistics of eigenvalues of Gaussian random matrices},
Phys. Rev. E {\bf 77}, 041108 (2008).

\bibitem{Dean16} D. S. Dean, P. Le Doussal, S. N. Majumdar and G. Schehr,
\emph{Statistics of the maximal distance and momentum in a trapped Fermi gas at low temperature},
arXiv:1612.03954.

\bibitem{Facchi10} A. De Pasquale, P. Facchi, G. Parisi, S. Pascazio and A. Scardicchio, 
\emph{Phase transitions and metastability in the distribution of the bipartite entanglement of a large quantum system},
Phys. Rev. A {\bf 81}, 052324 (2010).

\bibitem{Douglas93} M. R. Douglas and V. A. Kazakov, 
\emph{Large-N phase transition in continuum QCD in two-dimensions}, 
Phys. Lett. B {\bf 319}, 219  (1993).

\bibitem{Dyson62} F.~J. Dyson, 
\emph{Statistical Theory of the Energy Levels of Complex Systems}, 
J. Math. Phys. {\bf 3}, 140 (1962); {\bf 3}, 157 (1962); {\bf 3}, 166 (1962); {\bf 3}, 1191 (1962); {\bf 3}, 1199 (1962).

\bibitem{Edelman15} A. Edelman and M. La Croix,
\emph{The Singular Values of the GUE (Less is More)},
Random Matrices: Theory Appl. {\bf 04}, 1550021 (2015).

 \bibitem{Facchi08} P. Facchi, U. Marzolino, G. Parisi, S. Pascazio and A. Scardicchio, 
\emph{Phase Transitions of Bipartite Entanglement},
Phys. Rev. Lett. {\bf 101}, 050502 (2008).


\bibitem{Forrester98} P.~J. Forrester,
\emph{Exact results for two-dimensional Coulomb systems},
Phys. Rep. {\bf 301}, 235-270 (1998).

\bibitem{FMS11}
P.~J. Forrester, S. N. Majumdar and G. Schehr, 
\emph{Non-intersecting Brownian walkers and Yang-Mills theory on the sphere},
Nucl. Phys. B {\bf 844}, 500 (2011).


\bibitem{FN12}
Y. V. Fyodorov and C. Nadal, 
\emph{Critical Behavior of the Number of Minima of a Random Landscape at the Glass Transition Point and the Tracy-Widom Distribution},
Phys. Rev. Lett. {\bf 109}, 167203 (2012).
 

\bibitem{Ginibre65} J. Ginibre,
\emph{Statistical ensembles of complex, quaternion, and real matrices}, 
J. Math. Phys. {\bf 6}, 440-449 (1965).


\bibitem{Gross80} D. J. Gross and E. Witten,
\emph{Possible Third Order Phase Transition in the Large-N Lattice Gauge Theory}, 
Phys. Rev. D {\bf 21}, 446  (1980).


\bibitem{Jancovici93} B. Jancovici, J.~L. Lebowitz and G. Manificat, 
\emph{Large Charge Fluctuations in Classical Coulomb Systems},
J. Stat. Phys. {\bf 72}, 3/4 (1993).


\bibitem{MIMO} P. Kazakopoulos, P. Mertikopoulos, A. L. Moustakas and G. Caire, 
\emph{Living at the Edge: A Large Deviations Approach to the Outage MIMO Capacity},
IEEE T. Inform. Theory {\bf 57}, 1984 (2011). 


\bibitem{Kunz74} H. Kunz,
\emph{The One-Dimensional Classical Electron Gas},
Ann. Phys. {\bf 85}, 303-335 (1974).

\bibitem{Leble15} T. Lebl\'e and S. Serfaty,
\emph{Large Deviation Principle for Empirical Fields of Log and Riesz Gases},
arXiv:1502.02970. 

\bibitem{LiebLoss}
E.~H. Lieb and M. Loss,
\emph{Analysis}, 2nd Edition, 
American Mathematical Society (2001).


\bibitem{Lieb75} E. H. Lieb and H. Narnhofer, 
\emph{The Thermodynamic Limit for Jellium},
J. Stat. Phys. {\bf 12}, 291 (1975). 

\bibitem{Johansson99} K. Johansson,
\emph{Shape Fluctuations and Random Matrices},
Commun. Math. Phys. {\bf 209}, 437-476 (2000).


\bibitem{Spohn99}   M. K.-H. Kiessling and   H. Spohn,
\emph{A Note on the Eigenvalue Density of Random Matrices},
Commun. Math. Phys. {\bf 199}, 683-695 (1999).


\bibitem{Majumdar14}
S.~N. Majumdar and G. Schehr,
\emph{Top eigenvalue of a random matrix: large deviations and third order phase transition}, 
J. Stat. Mech.: Th. and Exp. {\bf P01012} (2014).

\bibitem{Majumdar09}
S.~N. Majumdar and M. Vergassola,
\emph{Large Deviations of the Maximum Eigenvalue for Wishart and Gaussian Random Matrices},
Phys. Rev. Lett. {\bf 102}, 060601 (2009).


\bibitem{Mehta04} M.~L. Mehta, 
\emph{Random Matrices}, 3rd Edition, 
Elsevier-Academic Press, (2004).

\bibitem{Nadal11} C. Nadal and S. N. Majumdar, 
\emph{A simple derivation of the Tracy-Widom distribution of the maximal eigenvalue of a Gaussian unitary random matrix}, 
J. Stat. Mech.: Th. and Exp. {\bf P04001} (2011).


\bibitem{NMV10}
C. Nadal, S. N. Majumdar and M. Vergassola, 
\emph{Phase Transitions in the Distribution of Bipartite Entanglement of a Random Pure State},
Phys. Rev. Lett. {\bf 104}, 110501 (2010). 

\bibitem{Rider03} B. Rider,
\emph{A limit theorem at the edge of a non-Hermitian random matrix ensemble},
J. Phys. A: Math. Gen. {\bf 36}, 3401-3409 (2003).


\bibitem{Rougerie15} N. Rougerie and S. Serfaty,
\emph{Higher-Dimensional Coulomb Gases and Renormalized Energy Functionals},
Comm. Pure Appl. Math.  {\bf 69}, 519-605 (2016).

\bibitem{SMCF}
G. Schehr, S. N. Majumdar, A. Comtet and P. J. Forrester, 
\emph{Reunion probability of N vicious walkers: typical and large fluctuations for large N},
J. Stat. Phys. {\bf 150}(3), 491 (2013).

\bibitem{VMB08}
P. Vivo, S. N. Majumdar and O. Bohigas, \emph{Distributions of Conductance and Shot Noise and Associated Phase Transitions}, Phys. Rev. Lett. {\bf 101}, 216809 (2008); \emph{Probability distributions of Linear Statistics in Chaotic Cavities and associated phase transitions}, Phys. Rev. B {\bf 81}, 104202 (2010). 



\bibitem{Wadia80} S. R. Wadia, 
\emph{$N = \infty$ Phase Transition in a Class of Exactly Soluble Model Lattice Gauge Theories}, 
Phys. Lett. B {\bf 93}, 403  (1980). 

\bibitem{Weigmann06} A. Zabrodin and P. Wiegmann,
\emph{Large-$N$ expansion for the 2D Dyson gas},
J. Phys. A: Math. Theor. {\bf 39}, 8933-8963 (2006).

\bibitem{Wigner34} E. Wigner,
\emph{On the Interaction of Electrons in Metals},
Phys. Rev. {\bf 46}, 1002 (1934).
\end{thebibliography}
\end{document}